\documentstyle[aps]{revtex}

\tightenlines

\begin{document}

\title{On Equivalence of Thermostatistical Formalisms }
\author{ G.A.
Raggio\thanks{Investigador CONICET; e-mail: raggio@fis.uncor.edu}}
\address{Facultad de Matem\'atica, Astronom\'\i a y F\'\i sica \\
Universidad Nacional de C\'ordoba \\
Ciudad Universitaria, 5000 C\'ordoba, Argentina}

\date{September, 1999}

\maketitle

\begin{abstract} 

We make some simple observations on basic issues pertaining to thermostatistical formalisms.

\end{abstract}

\vspace{.3in}

\section{Introduction}

This note proposes to comment on various issues connected with so-called thermostatistical formalisms. The main motivation is to put into  perspective some of the questions raised by three such formalisms put forth by C. Tsallis and co-workers \cite{T,C-T,TMP}.

\section{What is a thermostatistical formalism?}

We will discuss the subject in the domain of quatum mechanics\footnote{The discussion in the domain of classical mechanics is similar. But the entropies to be introduced in the following, also assume negative values.}. A state is associated with a positive, trace-class operator of unit trace (denoted by $tr$), i.e. a density operator.

Suppose $S[ \cdot ]$ is a function on the states  of a quantum
system taking non-negative real values (possibly including $+\infty$) such that
$S[ \cdot ]$ is {\em concave}: $S[ \lambda \rho_1 + (1-\lambda ) \rho_2 ] \geq \lambda S[\rho_1] + (1-\lambda ) S[\rho_2]$, for states $\rho_1 , \rho_2$ and $\lambda$ in the unit interval. Such a state-functional will be called and entropy.  In fact, in order to get
somewhere, one should demand that the set of states where $S[ \cdot
]$ takes finite values be {\em convex}, and that $S[ \cdot ]$ is
{\em strictly}
concave on this set, i.e. the inequality is strict whenever $\rho_1\neq \rho_2$,  $0 < \lambda < 1$ and the entropies involved are finite. This last property is shared by the following entropies:
\begin{enumerate}
\item The Boltzmann-Gibbs-von Neumann-Shannon entropy: $ S_1[\rho ]=- tr ( \rho \ln (\rho ))$.
\item The Havrda-Charvat/Dar\'oczy/Vajda/Tsallis ( \cite{H-C,D,T})\footnote{The reference to the work of Vajda is not available to me.} entropies or $q$-entropies: $S_q[\rho] = ( 1- tr( \rho^q))/(q-1)$ where $0 < q \neq 1$\footnote{See \cite{P}, for a collection of properties and the proof that these entropies satisfy the strict concavity property.}.  $S_q[ \cdot ]$ is also definable for $q\leq 0$, but we disregard these alltogether (see \cite{P}).
\item The R\'enyi entropies $S^R_q = (1- q)^{-1} \ln \left( tr ( \rho^q ) \right)$, where $0 < q < 1$\footnote{See \cite{Wehrl}.}. These are also defined for $q >1$ but are then not concave. 
\end{enumerate}
These three entropies also have the convenient feature that $S[\rho ]=0$ when and only when the state $\rho$ is pure. Moreover both the q-entropies and the R\'enyi entropies yield $S_1 [ \cdot ]$ when $q \to 1$.

Suppose further, that you have a certain ``expectation-value functional'' asssociating to each observable (i.e., operator) $A$ and each state $\rho$ a certain number $A[ \rho]$. If now you single out a particular observable, say the ``energy'' $H$, you are  in the position to set up a thermostatistical formalisms as follows.

Fixing a possible value $u$ of the ``energy'' functional $H[ \cdot]$ consider the variational problem
\begin{equation} \sup_{ \rho \,;\; H[\rho ]= u }  S[ \rho ]  \; =: S(u) \;,\label{Svar}
\end{equation}
which then, as one varies $u$, defines the ``entropy as a function of energy'' $S(u)$\footnote{We assume henceforth that the possible values of $u$ form an interval. Although the assumption is not necessary, it holds in the concrete formalisms mentioned later.}. We will systematically use square brackets $[\cdot ]$ to denote state-functionals to be distinguished from functions that will be associated with these functionals (e.g. $S(u)$). The questions that arise inmediately are, for example: Is the supremum in (\ref{Svar}) assumed? That is, are there states $\omega$ with $H[\omega ]=u$, $S[\omega ] = S(u)$? If so, is such a state unique?

The direct solution of a constrained variational problem is usually difficult. One proceeds to consider the Legendre-Fenchel transform of $S( \cdot )$ \begin{equation}
\phi ( \beta ) =  \inf_{u\,:\; u \;\;possible} \{ \beta u - S (u )\}  \;, \label{freie}
\end{equation}
for real $\beta$.  
By its very definition and irrespective of whether
$S( \cdot )$ is concave or not, the map $\beta \mapsto \phi ( \beta
)$ takes on extended real values (real values and $\pm \infty$), and is 
{\em concave} and {\em upper semicontinuos}. The function $\beta^{-1} \phi ( \beta )$ is the analogue
of the Helmholtz free energy (as a function of the reciprocal
``temperature'' $\beta$). Furthermore,
\begin{equation}
\phi ( \beta ) = \inf_{\rho}  \{ \beta H [ \rho ] - S [ \rho]  \} \label{freiefunk}
\end{equation}
since the infimum over the states  can be taken by
varying first over the states with $H[\rho ] =u$ and then over $u$. It is this  unconstrained variational problem which will eventually allow one to obtain  a solution of (\ref{Svar}).

Denoting by $\phi^* ( \cdot )$ the Legendre-Fenchel
transform of $\phi (\cdot )$ given by
\begin{equation}
\phi^* ( x) = \inf \{ x \beta - \phi ( \beta ) : \; \beta \in {\bf R}
\} \label{LFphi}
\end{equation}
for real $x$, we have, by a general result of the theory of
Legendre-Fenchel transforms (see e.g. \cite{Rocka}), that $\phi^*$ is
the least concave uppersemicontinuous function above $S(\cdot )$: 
\begin{equation}
S ( u) \; \leq \; \phi^* (u) \; ,  \label{Ungl}
\end{equation}
with equality for all possible $u$ if and only if $S( \cdot )$ is a {\em
concave, uppersemicontinuous} function. When this is the case, i.e. if one can  show that $\phi^* (\cdot ) = S( \cdot )$, and only in this case, one has the full-fledged Legendre transform structure familiar from Thermodynamics with all its consequences.

The basic relation between the constrained and unconstrained variational problems is given by\\

\noindent{\bf Lemma:} If for some $\beta_o$ the state $\rho_o$ is a  minimizer of (\ref{freiefunk}), i.e., $\phi( \beta_o)=\beta_oH[\rho_o ]- S[\rho_o] $, then $\rho_o$ is a maximizer of the problem (\ref{Svar}) for $u=u_o=H[\rho_o ]$ and   $S(u_o)=\phi^*(u_o)=S[\rho_o]$.\\

\noindent Proof: For any state $\rho$ with $H[ \rho ] = H[\rho_o]=u_o$,
\[ S[\rho ] =  \beta_o
u_o - \left( \beta_o H[\rho] - S [ \rho ] \right) \leq \beta_ou_o
- \phi ( \beta_o ) = S [ \rho_o ] \; ; \]
hence $\rho_o$ is a maximizer of (\ref{Svar}) and $S(u_o)= S[ \rho_o
]$. Moreover, due to (\ref{LFphi}),
\[ \phi^* (u_o ) \leq \beta_ou_o - \phi ( \beta_o ) = S [ \rho_o ] \;
, \]
so  (\ref{Ungl}) implies that $S(u_o)= \phi^* (u_o )$. Q.E.D.
\\

Now, the following things could --generally speaking-- go wrong when attacking the solution of (\ref{freiefunk}):
\begin{description}
\item [No Minimizers:] For a given $\beta_o$ there are no minimizers  of (\ref{freiefunk}), i.e. for every state $\rho$ one has $\phi ( \beta_o ) < \beta_o H[ \rho ] - S [ \rho ]$.
\item [Many Minimizers:] For a given $\beta_o$ there are two or more minimizers of (\ref{freiefunk})\footnote{This cannot happen if $H [ \cdot ]$ is affine and $S[ \cdot ]$ is
strictly concave because then the functional $\rho \mapsto \beta H [
\rho ] - S [ \rho ]$ is strictly convex.}. Consider the case of two
minimizing states $\rho_1$ and $\rho_2$, and let $u_j = H [ \rho_j
]$. The Lemma   gives
$S(u_j)=S[ \rho_j]$, and if $u_1=u_2$ all is well. If however $u_1
\neq u_2$ (and thus $S[ \rho_1 ] \neq S[ \rho_2]$) we cannot use
$\beta$ as a tag; $\beta \to H[ \mbox{ minimizers} ]$ is not a map. 
\item [No ``Temperature'':] For a given value $u_o $ one has $H[ \rho ]\neq 
u_o$ for all minimizers $\rho$  of (\ref{freiefunk}) for all real $\beta$; 
i.e.
$u_o$ never appears as the ``energy expectation-value'' of the states minimizing the
``free-energy'' functional $\rho \mapsto \beta H [ \rho ] - S[ \rho
]$.
\end{description}
One should thus like to address (and solve) the following points:
\begin{description}
\item [Unique Minimizers:] Find the set ${\cal B}$ of values of $\beta$ such
that $\phi ( \beta ) = \beta H[ \rho_{\beta } ] - S [ \rho_{\beta }
]$ for a {\em unique} state $\rho_{\beta}$; i.e. (\ref{freiefunk}) admits a
{\em unique} minimizer $\rho_{\beta}$.
\item [All ``Energies'' are ``Thermal'':] Does the range of the map ${\cal B} \ni \beta \mapsto H [ \rho_{\beta} ]$ coincide with the range of the ``energy functional''  $H[\cdot ]$, that is, the possible values of $u$? In other words, for every possible
energy $u$ is there is a $\beta$ such that $u = H[
\rho_{\beta}]$?
\item [One ``Temperature'' per ``Energy'':] Is the map ${\cal B} \ni \beta \mapsto H [ \rho_{\beta} ]$  invertible?
\end{description}

The completion of the above programme, that is to say the description of the set ${\cal B}$ and the positive answer to both questions, establishes -- via the Lemma -- the following  thermostatistical formalism:
\begin{enumerate}
\item $u \mapsto S(u)$ is concave and the Legendre-Fenchel transform of ${\cal B} \ni \beta \mapsto \phi( \beta )$.
\item for each possible value $u$ of the functional $H[\cdot ]$, there is a unique state $\omega_u$ with $S(u)=S[\omega_u]$ and a unique $\beta (u)$ such that $\omega_u=\rho_{\beta (u)}$; this  $\beta (u)  $ is  the inverse of the map ${\cal B} \mapsto H[\rho_{\beta }]$.
\end{enumerate}
The pair $S(\cdot )$, $\phi ( \cdot )$ then satisfies the same relations as in classical thermodynamics; further, the (intensive) parameter $\beta$ plays a role analogous to the reciprocal temperature; moreover, one has variational principles analogous to those of classical tehrmodynamics. Thus, one is tempted to interpret the maximizer $\omega_u$ of the problem (\ref{Svar}) and the minimizer $\rho_{\beta}$ of the problem (\ref{freiefunk}) as ``equilibrium'' states. 

\section{Examples \& Equivalence}

There are as many thermostatistical formalisms as distinct pairs of ``energy''/entropy functionals. However, apart from the considerations of the problems discussed at the end of the previous section which can often be solved in abstracto, in only a few of these formalisms one is able to obtain a manageable expression for $\rho_{\beta}$. When the underlying Hilbert space is infinite dimensional, one cannot expect a finite value for $\phi ( \beta )$ unless one imposes conditions on the spectrum of the operator $H$. The prime examples where the programme of the previous section can be carried out, are:
\begin{description}
\item [The standard formalism:] Uses the Boltzmann-Gibbs-von Neumann-Shannon entropy $S_1 [ \cdot ]$ and the standard quantum mechanical expectation-value functional
\begin{equation}
A^{(1)}[\rho] = tr (A \rho ) \;.\label{ewert}
\end{equation}
\item [The first Tsallis formalisms (\cite{T}):] Uses the q-entropy  $S_q[\cdot ]$, where $0< q \neq 1$, and the standard quantum mechanical  expectation-value functional (\ref{ewert}).
\item [The second Tsallis formalism (\cite{C-T}):] Uses the q-entropy $S_q[\cdot ]$, where $0< q \neq 1$, and the ``expectation-value functional''
\begin{equation}
A_q^{(2)}[\rho ] = tr ( A \rho^q )\label{2ewert}
\end{equation}
with the same value of $q$ as that used for the entropy.
\item [The third Tsallis formalism (\cite{TMP}):] Uses the q-entropy $S_q[\cdot ]$, where $0< q \neq 1$, and the ``expectation-value functional''
\begin{equation}
A_q^{(3)}[\rho ] = \frac{tr ( A \rho^q )}{tr ( \rho^q )} \label{3ewert}
\end{equation}
with the same value of $q$ as that used for the entropy.
\end{description}

Of the three above ``expectation-value functionals'' only the standard one has both of the following  properties: {\em linearity in the observables},
\begin{equation} (A+B)[\rho ] = A[\rho ] + B[\rho ]\;; \label{lin}
\end{equation}
and {\em ``linearity in the state''}, more precisely $\rho \mapsto A[\rho]$ is affine,
\begin{equation} 
A [ \lambda \rho_1 +(1-\lambda ) \rho_2 ] = \lambda A[\rho_1 ] +(1- \lambda ) A [ \rho_2] \;.\label{aff}
\end{equation}
(\ref{2ewert}) does not satisfy (\ref{lin}) and  (\ref{aff}), while for (\ref{3ewert}) one has (\ref{lin}) but not (\ref{aff}).
Both of these properties are crucial for a consistent probabilistic interpretation.

What about the three  formalisms based on the R\'enyi entropy and the three above ``expectation-value functionals''?  It was remarked in \cite{Ram} that 
 the R\'enyi entropies and the q-entropies are, for given $q$ and $\rho$, monotone increasing functions of each other. It then follows that the three possible formalisms based on R\'enyi entropies are in fact equivalent with the corresponding Tsallis' formalisms in the following sense: {\sl For fixed ``expectation-value functional'' $H[\cdot ]$ and fixed $q$, the formalism based on $S_q^R[\cdot ]$ predicts the same maximizer $\omega_u$ for the problem (\ref{Svar}) as the formalism based on $S_q[\cdot ]$}. Obviously, the associated entropy functions ($S(\cdot )$), ``free energy'' functions ($\phi ( \cdot )$) as well as the ``reciprocal temperature functions ($\beta (u)$), etc.,  are distinct. Thus, unless one has a reason to prefer one of the ensuing ``thermodynamics'' over the other, the formalisms are equivalent since they predict the same state and ``expectation-values'' for all observables.

To elaborate on this, consider an entropy-functional $S[\cdot ]$, let $X=\{ S[\rho ] :\; \rho \mbox{ a state}\}$ be the range of values of $S[\cdot ]$ and suppose $X$ is an interval (in $(0, \infty )$). If then a real-valued function $g$, defined on $X$, is monotone increasing then, obviously, the problem (\ref{Svar}) has the same maximizers for $S[\cdot ]$ as for $(g \circ S)[\cdot ]$ when $H[\cdot ]$ is fixed. Thus, in general, {\sl if $g$ is a monotone increasing function defined on the range of the entropy-functional $S[\cdot ]$, then the formalisms $(H[\cdot ], S[ \cdot ])$ and $(H[\cdot ] , (g\circ S)[\cdot ])$ are equivalent}. 
Moreover, if $g$ is concave  then so is $(g \circ S)[\cdot ]$ and this concavity is strict if both $S[\cdot ]$ and $g$ are strictly concave\footnote{For example: $ 0 \leq x \mapsto g_r(x)=x^{r}$ with $0 < r < 1$ are increasing strictly concave functions; they also have the convenient feature that $g_r(0)=0$.}. 

Consider (\cite{Ram})
\begin{equation} g_q (x) = (1-q)^{-1} \ln \left( 1 + (1-q)x \right) \;\;, x \geq 0 \;,\label{RT} \end{equation}
for $0 < q < 1$. Then $g_q$ is increasing, and strictly concave and $S_q^R[\rho ]= (g_q \circ S_q)[\rho ]=g_q ( S_q [\rho ])$.  For $q> 1$, (\ref{RT}) is defined for $0 \leq x < 1/(q-1)$, which coincides with the range of $S_q[\cdot ]$, and the corresponding function $g_q$ is still increasing but no longer concave; again $S_q^R [\rho ] = (g_q \circ S_q)[\rho]$. This substantiates the claims made above about equivalence of the R\'enyi and Tsallis' formalisms for the same ``expectation-value functional''.

Another, slightly less obvious example of equivalence was pointed out in \cite{R}, where  the formalisms $( A^{(1)} [\cdot ] , S_q [\cdot ])$ and $(A_{1/q}^{(3)} [\cdot ], S_{1/q} [ \cdot ])$ are shown to be equivalent in the same sense: they give the same ``expectation-values'' for all observables.

In the spirit exposed in the previous section, the first and second Tsallis' formalisms were studied in \cite{GR} and \cite{GPR} respectively. The three problems ``Unique Minimizers'', ``All ``Energies'' are ``Thermal'''', and ``One ``Temperature'' per ``Energy'''' were solved; but the solutions turn out to be positive only for the first Tsallis formalism. In the  second Tsallis formalism and depending on the spectrum of $H$, one can have no minimizers or many minimizers of problem (\ref{freiefunk}), and also ``Non-thermal energies''.

\section{Additivity}

First  some notation and well known facts. Consider two quantum systems described by Hilbert spaces ${\cal H}_1$ and ${\cal H}_2$ respectively. The composite system is described by the Hilbert space tensor product ${\cal H}={\cal H}_1 \otimes {\cal H}_2$. For any state $\rho$ of the compositum, $tr_{\cal H} (\rho (A_1\otimes I_2))$ and $tr_{\cal H} (\rho (I_1 \otimes A_2))$, $I_j$ denoting the identity operator on the respective Hilbert space, define states $\rho_1$ and $\rho_2$ of the subsystem 1 and 2 respectively. A state $\rho$ is called a product-state if
\[ tr_{\cal H} (\rho (A_1 \otimes A_2))= tr_{{\cal H}_1} ( \rho_1 A_1)tr_{{\cal H}_2} ( \rho_2 A_2) \]
and this is written as $\rho=\rho_1 \otimes \rho_2$. For such states there are no correlations whatsoever between observables associated to the different subsystems.  If the Hamiltonian governing the time evolution of the compositum has interaction-terms between the subsystems, then the evolution of a product-state is no longer a product-state. This feature is there irrespective of the purity of the initial product-state\footnote{A product-state $\rho_1\otimes \rho_2$ is pure iff both $\rho_1$ and $\rho_2$ are pure.}, and is one of the distinctive properties of quantum mechanics as compared to classical mechanics.

We are very familiar with two properties of the standard formalism. First, entropy is additive over ``independent'' subsystems:
\begin{equation}
S_1 [ \rho_1 \otimes \rho_2 ] = S_1 [ \rho_1] + S_1 [ \rho_2] \label{additivitaet}\;.
\end{equation}
And, secondly, the Gibbs state (i.e. the solution of the problem (\ref{freiefunk}) in the standard formalism) of a non-interacting compositum is the product-state  of the Gibbs states of the subsystems:
\begin{equation}
 \Phi_{\beta} =\frac{\exp\{ -\beta ( H_1 \otimes I_2 + I_1 \otimes H_2 )\}}{tr_{\cal H} (\exp \{ -\beta ( H_1 \otimes I_2 + I_1 \otimes H_2 )\})} = \left( \frac{\exp \{ -\beta  H_1\}}{tr_{{\cal H}_1} (\exp \{ -\beta  H_1\} )}\right) \otimes \left( \frac{\exp \{ -\beta  H_2\}}{tr_{{\cal H}_2} (\exp \{ -\beta  H_2\} )}\right) = \Phi_{\beta}^{(1)} \otimes \Phi_{\beta }^{(2)}\;\;.\label{Gibbs}
\end{equation}
This simple, well-known, yet very remarkable property of the standard formalism is responsible for the use of the reciprocal temperature $\beta$ as a tag or label for the transitive relation of thermal equilibrium.

Moreover, since the solution of the problem (\ref{Svar}) is the Gibbs-state $\Phi_{\beta (u)}$ at reciprocal temperature $\beta (u)$ where $\beta ( \cdot )$ is the inverse of the decreasing function\footnote{${\cal B}= (-\infty, \infty )$ for finite systems; for infinite systems with Hamiltonian $H$ bounded below and a condition on the (purely discrete) spectrum guaranteeing existence of $\Phi_{\beta }$, one has ${\cal B}=(0, \infty )$.}  ${\cal B}\ni \beta \mapsto (H_1\otimes I_2+I_1\otimes H_2)[\Phi_{\beta}]$, one has, due to (\ref{lin}) and (\ref{additivitaet}), the property
\begin{equation} S(u) = \sup \{ S(u_1)+ S(u_2) \}\;,\label{add}
\end{equation}
where the supremum is taken over the possible energies $u_1$ of subsystem 1, and the   possible energies $u_2$  of subsystem 2, with $u_1+u_2=u$.

It is inmediately verified that the R\'enyi entropies are additive whereas the q-entropies are not.\\

Obviously (\ref{add}) is directly related to additivty (\ref{additivitaet}). But the reason behind the identity (\ref{Gibbs})  is not directly related to additivity (although this property is an experimental fact)! Indeed, for any increasing and strictly concave function with the correct domain of definition, the equivalent formalism $((H_1\otimes I_2+I_1\otimes H_2)[\cdot ], (g \circ S_1)[\cdot ])$ predicts the same maximizer of (\ref{Svar}) although, of course, the associated map $\beta (\cdot )$ is different. But $(g \circ S_1)[\cdot]$ need not be additive, e.g., $ \rho \to \sqrt{S_1[\rho ]}$ does not satisfy (\ref{additivitaet}).

Conversely, (\ref{additivitaet}) does not necessarily imply the product-property characteristic of the Gibbs-state (\ref{Gibbs}). Indeed, $S_q^{R} [ \cdot ]$ satisfies (\ref{additivitaet}) but the formalisms based on  $S_q^{R} [ \cdot ]$, which are equivalent to those based on $S_q[\cdot ]$, do not have this property (see the formulas for $\rho_{\beta}$ in the original references \cite{T,C-T,TMP}, and in \cite{GR,GPR}).

\end{document}